\documentclass[twocolumn,aps,pra,showpacs,raggedbottom,nobalancelastpage,amssymb,superscriptaddress,longbibliography]{revtex4-1}

\usepackage{graphicx}
\usepackage{amsmath}
\usepackage{amssymb}
\usepackage{bm}
\usepackage[usenames]{color}
\usepackage{amsfonts}

\usepackage{comment}

\usepackage[colorlinks=true , citecolor=blue,urlcolor=blue]{hyperref}

\newcommand{\ie}{{i.e.}}

\def\kb{{\mathchar'26\mkern-9mu k}}

\usepackage{ulem}

\include{./journalsVaps}

\begin{document}
\title{Dynamics of the mean-field interacting quantum kicked rotor}
\author{Samuel Lellouch}
\affiliation{Univ. Lille, CNRS, UMR 8523 -- PhLAM -- Physique des Lasers Atomes et Mol\'ecules, F-59000 Lille, France}
\author{Adam Ran\c con}
\affiliation{Univ. Lille, CNRS, UMR 8523 -- PhLAM -- Physique des Lasers Atomes et Mol\'ecules, F-59000 Lille, France}
\author{Stephan De Bi\`evre}
\affiliation{Univ. Lille, CNRS, UMR 8524, Inria - Laboratoire Paul Painlev\'e, F-59000 Lille, France}
\author{Dominique Delande}
\affiliation{Laboratoire Kastler Brossel, UPMC-Sorbonne Universit\'e, CNRS, ENS-PSL Research University, Coll\`ege de France, 4 Place Jussieu, 75005 Paris, France}
\author{Jean Claude Garreau}
\affiliation{Univ. Lille, CNRS, UMR 8523 -- PhLAM -- Physique des Lasers Atomes et Mol\'ecules, F-59000 Lille, France}

\begin{abstract}
We study the dynamics of the many-body atomic kicked rotor with interactions at the mean-field level, governed by the Gross-Pitaevskii equation. We show that dynamical localization is destroyed by the interaction, and replaced by a subdiffusive behavior. In contrast to results previously obtained from a simplified version of the Gross-Pitaevskii equation, the subdiffusive exponent does not appear to be universal. By studying the phase of the mean-field wave function, we propose a new approximation that describes correctly the dynamics at experimentally relevant times close to the start of subdiffusion, while preserving the reduced computational cost of the former approximation.
\end{abstract}

\date{\today}
\maketitle
%%%%%%%%%%%%%%%%%%%%%%%%%%%%%%%%%%%%%%%%%%%%%%%%%%%%%%%%%%%%%%%%%%

\section{Introduction}

Ultracold quantum gases are versatile tools to simulate a great variety of condensed-matter systems~\cite{Bloch:ManyBodyUltracold:RMP08}. In many cases, the use of \textit{equivalent models} -- \ie~alternative models that can be mapped onto a desired system -- constitutes a promising route, at lower experimental or computational expense, to a better understanding of the underlying physics. A well-known illustration is Floquet engineering, which relies on the careful design of time-periodic systems whose stroboscopic evolution is governed by an effective static Hamiltonian featuring the desired properties~\cite{Goldman:GaugeFieldsUltracoldAtoms:RPP14,Eckardt:HighFreqApproxFloquet:NJP15}. 
Floquet engineering allows mimicking to a high degree of details the physics of quantum disordered systems. In this context, the atom-optics realization~\cite{Moore:AtomOpticsRealizationQKR:PRL95} of the quantum kicked rotor (QKR)~\cite{Chirikov:ChaosClassKR:PhysRep79,Casati:LocDynFirst:LNP79} has proven to be an almost ideal quantum simulator~\cite{Hainaut:IdealFloquetSystem:NJP19}. 
The associated dynamics is known to display dynamical localization, which is the analog of Anderson localization in momentum space~\cite{Anderson:LocAnderson:PR58}, and the QKR  can, in fact, be rigorously mapped onto a 1D Anderson model~\cite{Fishman:LocDynAnders:PRL82}. Moreover, by adding to the kick amplitude a temporal dependence made of $d-1$ frequencies incommensurate with the kick frequency $T_1^{-1}$, one obtains the quasiperiodic kicked rotor which maps to an Anderson model in $d$ dimensions~\cite{Casati:IncommFreqsQKR:PRL89,Chabe:Anderson:PRL08,Lemarie:AndersonLong:PRA09}.
Thanks to its experimental and conceptual simplicity, the QKR has been widely used to investigate Anderson-like physics  experimentally: observation of Anderson transition~\cite{Chabe:Anderson:PRL08}, characterization of critical properties~\cite{Lemarie:AndersonLong:PRA09,Lemarie:CriticalStateAndersonTransition:PRL10,
Lopez:ExperimentalTestOfUniversality:PRL12,Lopez:PhaseDiagramAndersonQKR:NJP13}, critical dimension localization~\cite{Manai:Anderson2DKR:PRL15} or other universality classes~\cite{Hainaut:CFS:NCM18}.

A challenging question concerns many-body effects on Anderson localization~\cite{Anderson:LocAnderson:RMP78,SanchezPalencia:DisorderQGases:NP10}. This question arises in a wide class of condensed-matter systems, from disordered supraconductors~\cite{Fiory:ElectronMobilityMetalInsulTrans:PRL84,Belitz:AndersonMottTrans:RMP94} to superfluid $^4$He in porous media~\cite{Crooker:SuperfluidityDiluteBoseGas:PRL83}. In addition, the interplay between disorder and interactions is known to give rise to non-trivial collective behavior, possibly  underlying complicated many-body phase transitions~\cite{Aleiner:1DPhaseTrans:NP10,Basko:MetalInsulatorMBL:ANP06,Lugan:AndersonLocalizationBogolyubov:PRL07,Lugan:LocalizationBogoliubovQuasip:PRA11,Cazalilla:1DBosonsCondMattUGases:RMP11,Lellouch:LocTransWeaklyInt1DQPLatt:PRA14,Lellouch:PropagationCollectivePairEx:PRA15,
Altman:SuperfluidInsulatorTransitionBosons1D:PRB10}. In the more restricted frame of bosonic systems in the mean-field interacting regime, where interactions are simply taken into account by a quadratic nonlinearity in the Gross-Pitaevskii Equation (GPE)~\cite{Dalfovo:BECRevTh:RMP99}, previous studies of the mean-field Anderson model have predicted that the localized regime should be replaced by a subdiffusive phase~\cite{Pikovsky:DestructionAndersonLocNonlin:PRL08,Flach:DisorderNonlin:PRL09,Laptyeva:DisorderNonlinChaos:EPL10,
Cherroret:AndersonNonlinearInteractions:PRL14}. In 3D however, the full numerical simulation of the mean-field Anderson model implies a very heavy computational cost. Hence, the QKR could constitute a promising equivalent model to circumvent this issue; yet, it is still unclear whether its equivalence with the corresponding Anderson model holds in the presence of interactions, since the latter are local in position space whereas localization occurs in momentum space. 

The mean-field interacting bosonic QKR is modeled by a GPE~\cite{Shepelyansky:KRNonlinear:PRL93}, see Eq.~\eqref{eq:GPE} below. 
The nonlinear term in GPE prompts the appearance of \emph{quasiclassical} chaotic behaviors, i.e. chaos related to a sensitivity to initial conditions, a subject that also attracted much interest in various kinds of degenerated quantum-gas systems, both kicked~\cite{Gardiner:NonlinearMatterWaveDynamics:PRA00,Filho:ChaosCollapsingBEC:PRA00,
Chiofalo:RoutesToChaosForDrivenBECs:PLA02,Trombettoni:SolitonsBEC:PRL01} 
and non-kicked~\cite{Thommen:ChaosBEC:PRL03,Smerzi:InstabilityBEC:PRL04,Lucioni:SubdiffusionInteract:PRL11,
Fallani:InstabilityBEC:PRL04,Weiss:DifferencesMeanFieldQuantumDynamics:PRL08,Kagan:CollapseBECNegScatLenght:PRL98,
Carvalho:DecoherencePhaseSpaceKHO:PRE04}. 
%For the interacting quantum kicked rotor, the dynamics described by the GPE has  been  studied  in  various  regimes  \cite{Shepelyansky:KRNonlinear:PRL93,Benvenuto:ManifestationsClassicalQuantumNonlinWaveProp:PRA91,Wimberger:QRScalingLaw:PRA05,Monteiro:QRBEC:PRL09,Gligoric:InteractionsDynLocQKR:EPL11,Vakulchyk:WavePacketSpreadingDisordered:PRL19,Shepelyansky:KolmogorovTurbulence:EPJB20}, but a systematic study of  the  interplay between the interactions and the kicking strength on the long-term dynamics of  the  exact GPE is to the best of our knowledge not available in the literature,  due  to  strong  numerical  instabilities  and computer  resource  cost.
For the interacting quantum kicked rotor, the dynamics described by the GPE has  been  studied  in  various  regimes  \cite{Shepelyansky:KRNonlinear:PRL93,Benvenuto:ManifestationsClassicalQuantumNonlinWaveProp:PRA91,Wimberger:QRScalingLaw:PRA05,Monteiro:QRBEC:PRL09,Gligoric:InteractionsDynLocQKR:EPL11,Vakulchyk:WavePacketSpreadingDisordered:PRL19,Shepelyansky:KolmogorovTurbulence:EPJB20}. However, a systematic study of  the effects of the interplay between the interactions and the kicking strength on the long-term dynamics of  the  exact GPE is thus far lacking due  to  strong  numerical  instabilities  and computer  resource  costs.

For these reasons, the long-time dynamics of the interacting QKR has instead mostly been studied with an uncontrolled approximation [that we dub the {\em local momentum approximation} (LMA), see below]~\cite{Shepelyansky:KRNonlinear:PRL93,Cherroret:AndersonNonlinearInteractions:PRL14,Flach:DisorderNonlin:PRL09,
Rebuzzini:KRNonlinQuantumRes:PRE05}. The latter predicts a subdiffusion of the kinetic energy $\langle p^2(t)\rangle\propto t^{\alpha}$ ($\alpha\sim 0.3 - 0.4$), thus implying a destruction of dynamical localization, which corresponds to $\alpha=0$. Furthermore, the exponent $\alpha$ is similar to that found in the 1D interacting Anderson model at mean-field level (the equivalence between QKR and Anderson model being preserved in the framework of the LMA), and is expected to be universal. 

There is thus a need to study the {\em exact} GPE of the interacting QKR, to ensure that this subdiffusion is indeed present and to test the universality of the subdiffusion exponent. In this work, we study the asymptotic behavior of the GPE at very long times (up to $10^5$ kicks), and show that, while our data is compatible with subdiffusion, they are not described by a universal exponent. We offer an improved approximate dynamics which is numerically shown to be a better approximation of the full GPE approximation for experimentally accessible times (about $10^3$ kicks).

\section{The nonlinear quantum kicked rotor}

We consider a degenerate boson gas in a ring of circumference $L_\parallel$, tightly confined in the transverse direction with a characteristic energy scale $\hbar \omega_{\perp}$ and transverse dimension $L_{\perp}\ll L_\parallel$, such that this energy is much larger than any other energy scale in the system. In such conditions the dynamics of the transverse degrees of freedom is effectively frozen and the longitudinal dynamics is essentially one dimensional.
The mean-field bosonic QKR wave function $\psi(x,t)$ is governed by the GPE
\begin{equation}
i\kb\partial_t\psi = -\kb^2 \frac{\partial_x^2\psi}{2} +g|\psi|^2\psi+K\cos(x)\sum_{n \in \mathbb{Z}} \ \delta(t-n)\psi,
\label{eq:GPE}
\end{equation}
where $K$ is the kick amplitude, proportional to the optical potential created by a pulsed standing wave of wavenumber $k_L$, time is expressed in units of the interval $T_1$ between two kicks, lengths are in units of $(2k_L)^ {-1}$ and the effective Planck constant is  $\kb=4\hbar k_L^2 T_1/M$ with $M$ the atom mass \cite{Lemarie:AndersonLong:PRA09}. The dimensionless 1D nonlinear coupling constant in such units is given by 
\begin{equation}
\label{eq:g}
g = \pi \kb^2 k_L a \frac{\hbar }{\omega_R M L_{\perp}^2}N = \frac{\kb^2}{2}  k_L a \frac{\omega_\perp}{\omega_R}N,
\end{equation}
where the last equality is valid for a harmonic transverse confinement, see for instance~\cite{Alexander:DynamicsMatterWaveSolitonsHarmonicTraps:PRA12}, $a$ is the (3D) $s$-wave scattering length of the contact interaction, $N$ the number of atoms, and $\omega_R = \hbar k_L^2/(2M)$ the recoil frequency. With typical values for  Potassium atoms, one obtains $g=1$ for $L_\perp = 5\, \mu\mathrm{m}$ (or, equivalently, $\omega_\perp/2\pi  = 62$ Hz) and $N \sim 1600$.
%%-----------------------------------------%
\begin{figure}[t!]
		\centering
		\includegraphics[width=1.02\columnwidth]{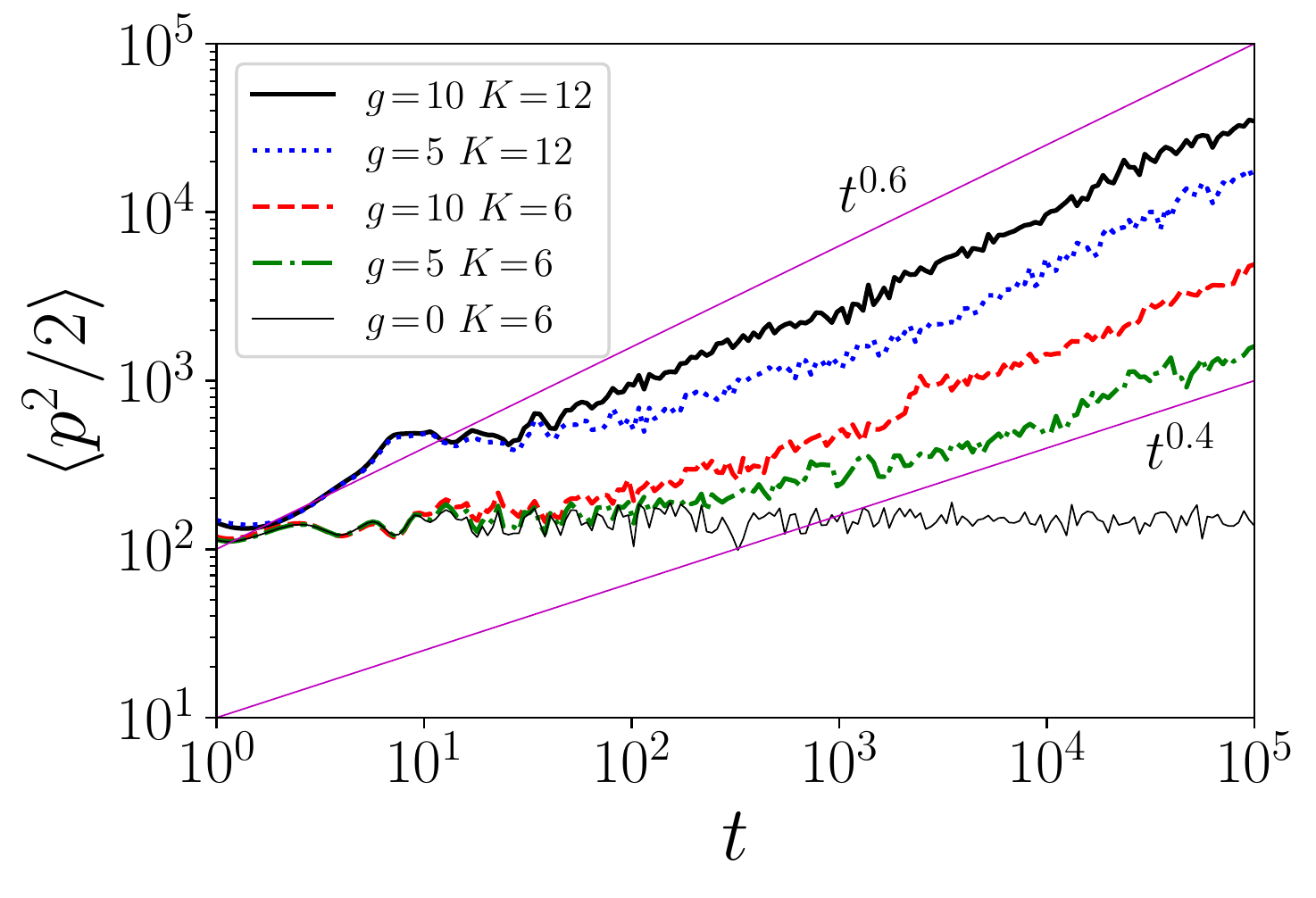}
		\caption{ 
		Evolution of the kinetic energy in the GPE QKR, for different combinations of $g$ and $K$,
 and with $\kb=2.89$. The long-time dynamics reveals a universal subdiffusion with a non-universal subdiffusive exponent -- the lines $t^{0.4}$ and $t^{0.6}$ are guides for the eye. The lowest curve is for the non-interacting kicked rotor $g=0$ and displays dynamical localization at long time.}
		\label{fig:GPEdyn}
\end{figure}
%-----------------------------------------%

We use periodic boundary conditions $\psi(L_\parallel,t)=\psi(0,t)$ where
$L_\parallel=2\pi$ is the system size and normalization
$\int_0^{L_\parallel} |\psi(x,t)|^2dx=1.$ Because of this boundary condition, the
spectrum of the momentum operator $p=i\kb \partial/\partial x$ is
discrete, $n\kb$ with $n$ an integer, so that the momentum space
wave function $\hat\psi(p)$ is given by the Fourier series:
\begin{equation}
\psi(x) = \frac{1}{\sqrt{L_\parallel}} \sum_n{\hat\psi(n\kb)\ \mathrm{e}^{inx}},
\end{equation}
with normalization $\sum_n{|\hat\psi(n\kb)|^2} = 1.$
The initial state is chosen to be delta-peaked in momentum space, $\hat \psi(p,t=0)=\delta_{p,p_0}$ and results are averaged over the initial momentum $p_0$.

We solve Eq.~(\ref{eq:GPE}) numerically by a real-time propagation. Two classes of unconditionally stable numerical methods can be used for this type of problems: a Crank-Nicolson scheme and a split-step method. Because of the high oscillations in space of the solution that occur as time evolves, both methods start to develop instabilities after some time, and one has to scale carefully the time step with the final time. Our numerical experiments indicate that such instabilities start developing earlier with the Crank-Nicolson method than with the split-step method. However, the split-step method also presents numerical instabilities, and long-time calculations demand higher-order split-step methods to be used. For all the results presented here, we used a second-order split-step method, with a time step $\Delta t=10^{-4}$, which prevents numerical instability in the time range $[0,10^6]$ (verified by studying the convergence as a function of the time step).
%
%
%
		%%-----------------------------------------%
\begin{figure}[t!]
		\centering
		\includegraphics[width=1.8\columnwidth]{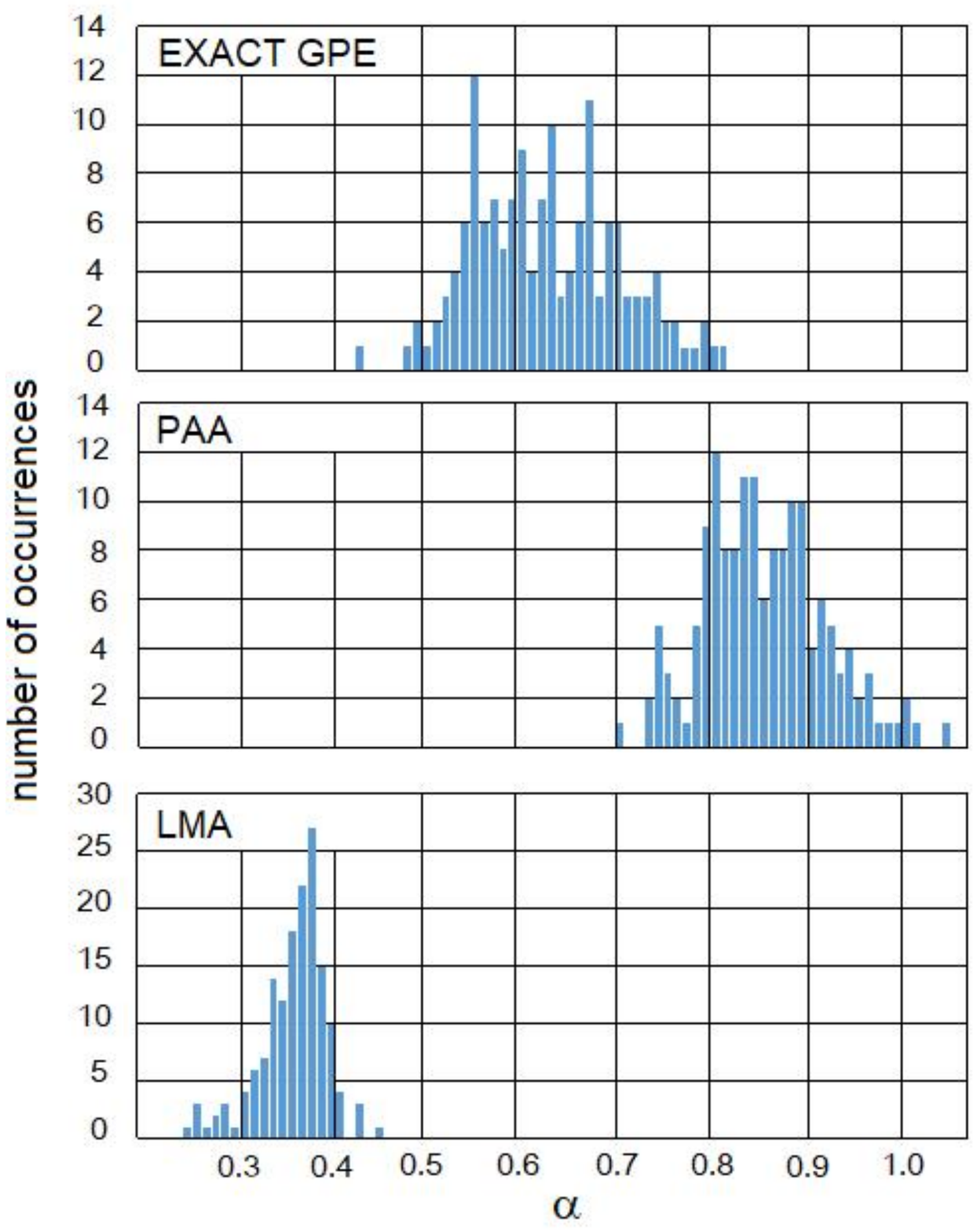}
		\caption{Statistics of the subdiffusive exponents found for the three models studied in this work (top panel: exact GPE, middle: PAA, bottom: LMA, see Sec.~\ref{sec:DAimproved}). Each histogram is obtained by varying the parameters $g$ and $K$ in $]0,20]$, and extracting for each parameter set a subdiffusive exponents by an algebraic fit of the kinetic energy vs. time -- i.e. a linear regression in a double logarithmic scale -- in the asymptotic regime ($t\in[10^4,10^5]$ for all models). The uncertainty on the exponent is about $0.1$ for each model. The dispersion in each 
set of results suggests a non-universality of the subdiffusive exponent. The disagreement between the LMA and PAA with the exact GPE shows their inability to capture correctly the asymptotic subdiffusive regime.}  
		\label{fig:hist} 
\end{figure}
%-----------------------------------------%
%
%
%
The resulting evolution of the kinetic energy is displayed in Fig.~\ref{fig:GPEdyn}, for various combinations of $g$ and $K$ (with $\kb=2.89$). The curves are averaged over 10 different choices of $p_0 \in [\kb, 10\kb]$. While the short-time dynamics is interaction-independent, and well described by the non-interacting QKR, the long-time one features subdiffusive behavior with a non-universal exponent $\alpha$. As observed in the top panel of Fig.~\ref{fig:hist}, this exponent generically falls in the range $[0.4-0.8]$ depending on the values of the parameters. We estimate the uncertainty of the numerically fitted exponents to be of order $0.1$ for each set of parameters, due to the choice of the time window of the fits (typically $t\in[10^4,10^5]$) and to the noise of the kinetic energy, which is smaller than the range of $\alpha$. There exist in the literature models predicting values for the subdiffusive exponent, relying on the nonlinearity-generated chaotic behavior, in both the nonlinear QKR and the nonlinear Anderson model~\cite{Shepelyansky:KRNonlinear:PRL93,Pikovsky:DestructionAndersonLocNonlin:PRL08,
Flach:DisorderNonlin:PRL09}. Ref.~\cite{Shepelyansky:KRNonlinear:PRL93} predicts $\alpha = 2/5$ with an argument based on the Chirikov resonance overlapping criterion~\cite{Chirikov:ChaosClassKR:PhysRep79}. Using a somewhat different argument based on weak chaos Ref.~\cite{Flach:DisorderNonlin:PRL09} found $\alpha = 1/3$. Due to the apparent non-universality of the sudiffusive exponent, both predictions can match numerical curves.

We thus confirm the existence of a subdiffusive behavior of the kinetic energy for the exact solution of the GPE, but could not confirm the universality of the corresponding exponent. We could not find any obvious pattern in the dependence of $\alpha$ on the parameters $\kb$, $K$ and $g$, and a better theoretical understanding of this dependence is beyond the scope of this work.

\section{Alternative models with approximated dynamics}
\label{sec:DAimproved}

In momentum space, the interaction term is nonlocal and given by the Fourier transform of $|\psi|^2\psi$, namely
\begin{equation}
F(p,\hat\psi)\equiv \frac{1}{2\pi}\sum_{p_1,p_2} \hat\psi^*(p_1)\hat\psi(p_2)\hat\psi(p+p_1-p_2) .
\label{eq:nonlocInt}
\end{equation}
In Ref.~\cite{Shepelyansky:KRNonlinear:PRL93}, a simplification was introduced neglecting all ``off-diagonal" contributions, thus making ``by hand'' the interaction local in momentum space,  which amounts to replacing it by 
\begin{equation}
F_{\rm LMA}(p,\hat\psi)=\frac{\gamma}{2\pi}|\hat\psi(p)|^2\hat\psi(p).
\label{eq:DA}
\end{equation}
where $\gamma$ could be a function of $p$ that cannot be easily determined, and that has been arbitrarily set to 1 in Ref.~\cite{Shepelyansky:KRNonlinear:PRL93}, a choice that we kept in this work.
Although not rigorously justified, this ``local momentum approximation'' (LMA) significantly reduces the cost of the numerical integration, since the resulting GPE can be integrated in momentum space between consecutive kicks. Moreover,  the simplified equation evokes a mean-field interacting Anderson model translated in the momentum space. 
A rationale for the LMA is to suppose that the evolution makes the phases of the different $\hat\psi(p,t)$ uncorrelated for large enough momentum differences, so that the integrals in Eq.~(\ref{eq:nonlocInt}) are dominated by the contributions of $p_1\simeq p_2\simeq p$. Finding a rigorous  derivation of the LMA is not obvious, but this heuristic argument can be used to construct an improved approximation, which we now describe.

%-----------------------------------------%
\begin{figure}[t!]
			\includegraphics[width=0.95\columnwidth]{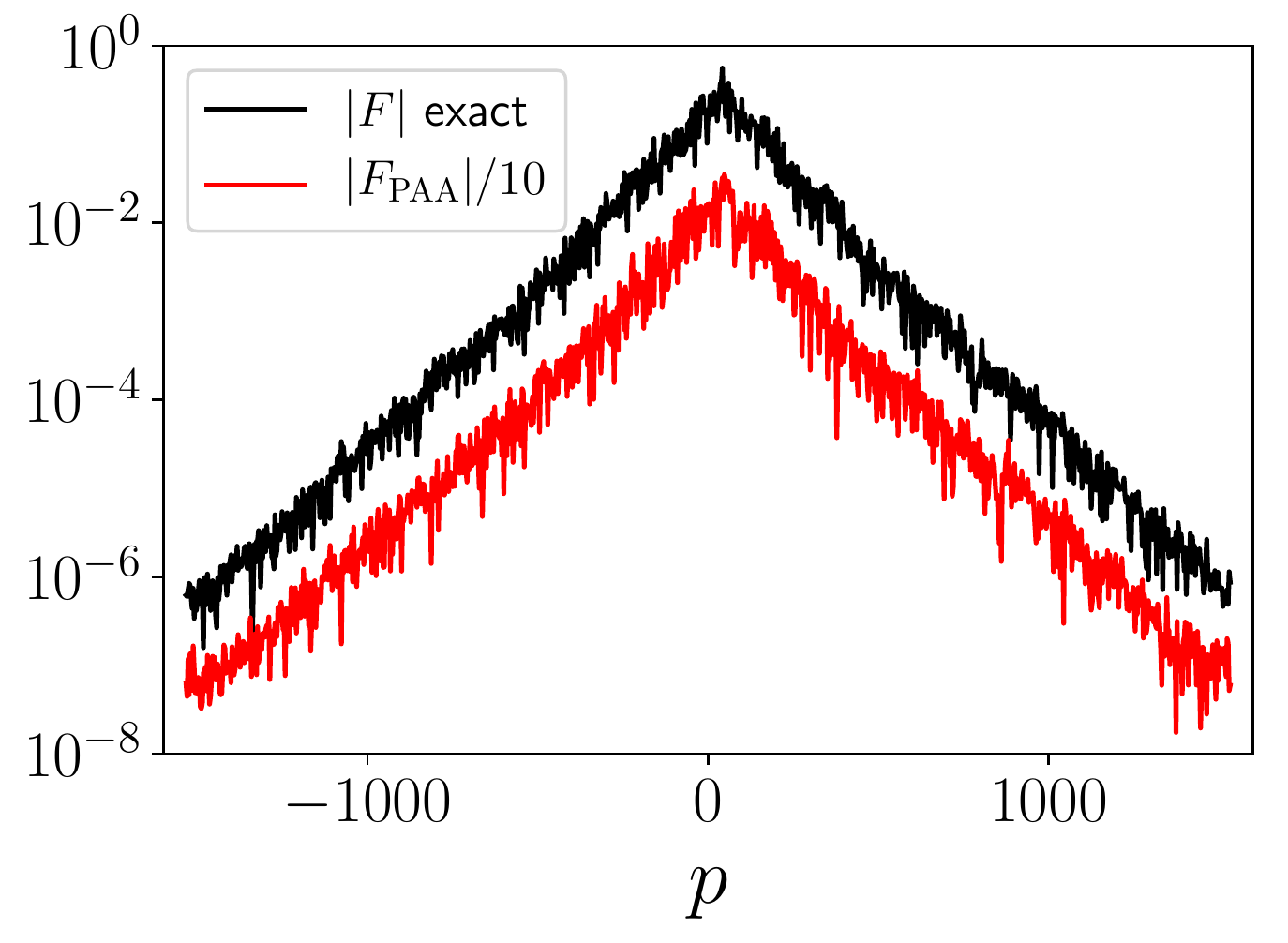}
			\caption{$|F(p,\hat\psi)|$, Eq.~(\ref{eq:nonlocInt}), computed from a solution $\hat \psi$ of the GPE QKR after $200$ kicks with parameters $K=12$, $g=12$, and $\kb=2.89$, and its approximation $|F_{\rm PAA}(p,\hat\psi)|$, on a logarithmic scale. The latter curve has been shifted downwards by one decade for clarity. The agreement is very good, validating the PAA, see Eq.~(\ref{eq:PAA}).}
			\label{fig:Fp_app}
		\end{figure}
%-----------------------------------------%

Writing the wave function in the amplitude-phase representation  $\hat\psi(p)=A(p) e^{i\phi(p)}$, our starting point is the numerical observation that in the presence of kicks, $\phi(p)$ obtained from the full GPE solution appears to be a uniformly randomly distributed function of $p$. In particular, we observe numerically that replacing the phases $\phi(p)$ by independent random phases distributed uniformly in $[0,2\pi[$, and thus $\hat\psi(p)\to\hat\psi_{\rm rand}(p)$, does not change the (modulus of the) interaction functional $|F(p,\hat\psi)|\simeq|F(p,\hat\psi_{\rm rand})|$.% We can thus treat the phase as a uniformly distributed random function of $p$.

Denoting the average over random $\phi(p)$ realizations by an overline, we have $\overline{F(p,\hat\psi)}=0$, because it only involves three uncorrelated phases. For the correlations $\overline{F(p,\hat\psi)F^*(p^\prime,\hat\psi)}$ we have
\begin{widetext}
\begin{equation}
\begin{split}
\overline{F(p,\hat\psi)F^*(p^\prime,\hat\psi)}=\frac{1}{(2\pi)^2}\sum_{p_1,p_2,p_1^\prime,p_2^\prime} &\,  A(p_1)A(p_2)A(p+p_1-p_2)A(p_1^\prime)A(p_2^\prime)A(p^\prime+p_1^\prime-p_2^\prime) \\
&\times \overline{\exp\left[-i(\phi(p_1)-\phi(p_2)-\phi(p+p_1-p_2)-\phi(p_1^\prime)+\phi(p_2^\prime)+\phi(p^\prime+p_1^\prime-p_2^\prime))\right]}.
\end{split}
\end{equation}
Using the independence of the phases, a straightforward calculation gives
\begin{equation}
\begin{split}
\overline{F(p,\hat\psi)F^*(p^\prime,\hat\psi)}=\frac{\delta_{p,p^\prime}}{(2\pi)^2}\left(4 A(p)^2+2 \sum_{p_1,p_2}\, A(p_1)^2A(p_2)^2A(p+p_1-p_2)^2\right).
\end{split}
\end{equation}
Furthermore, a numerical study of the phase of $F(p,\hat\psi)$ shows that it also appears to be uniformly distributed, and a calculation similar to that above shows that $\overline{F(p,\hat\psi)\hat\psi^*(p^\prime)}=\frac{\delta_{p,p^\prime}}\pi A(p)^2$. Therefore, the phase of $F(p,\hat\psi)$ is strongly correlated to that of $\hat\psi(p)$, and can thus be replaced by $\phi(p)$.

This suggests the following ``phase-averaging approximation'' (PAA) of $F(p,\hat\psi)$,
\begin{equation}
F_{\rm PAA}(p,\hat\psi)=\frac{e^{i\phi(p)}}{2\pi}\sqrt{4 A(p)^2+2 \sum_{p_1,p_2}\, A(p_1)^2A(p_2)^2A(p+p_1-p_2)^2}.
\label{eq:PAA}
\end{equation}
\end{widetext}
The agreement between $|F_{\rm PAA}(p,\hat\psi)|$ and the $|F(p,\hat\psi)|$ for a typical $\hat \psi$ is very good, as shown in Fig.~\ref{fig:Fp_app}.

%-----------------------------------------%
\begin{figure}[t!]
		\centering
		\includegraphics[width=\columnwidth]{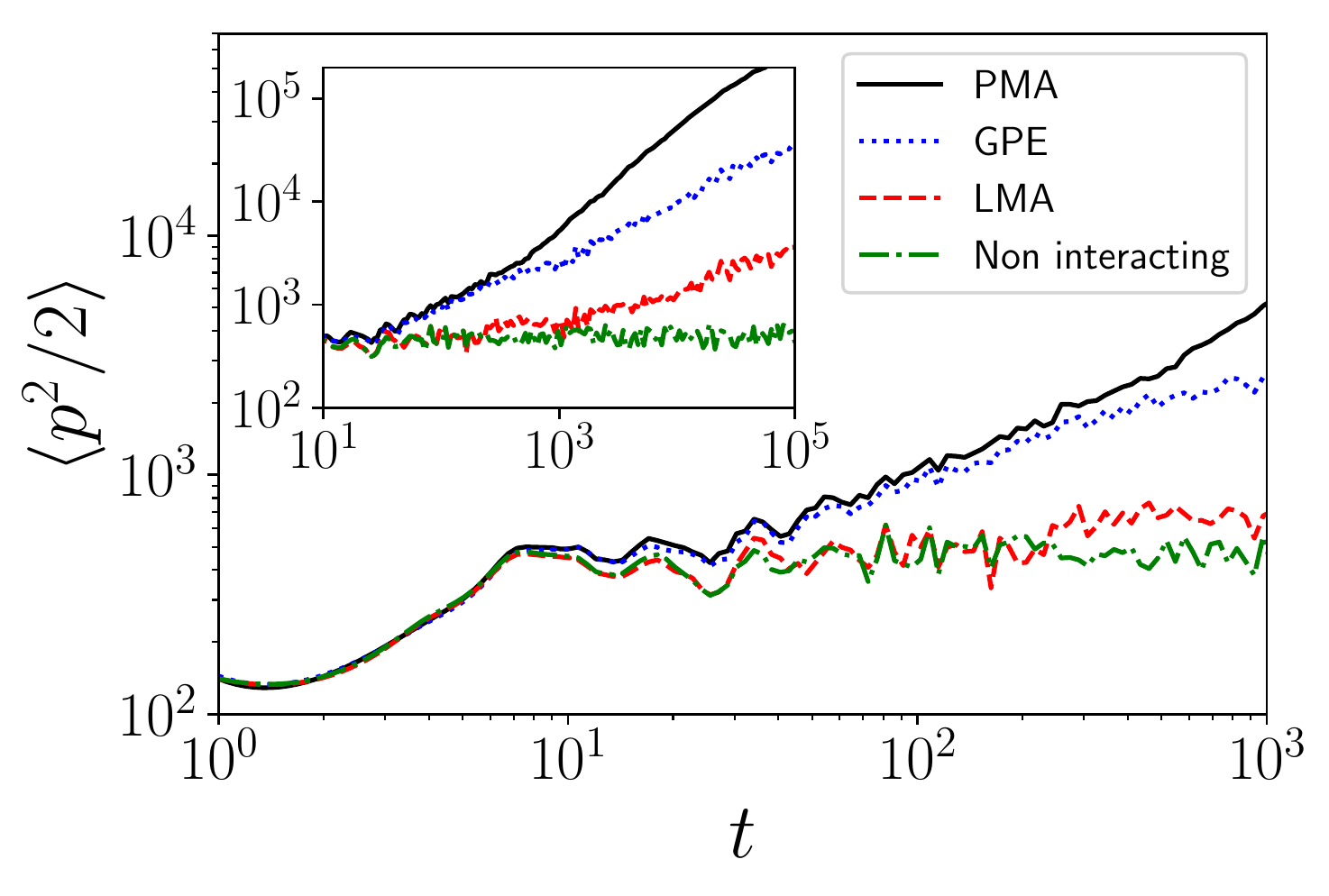}
		\caption{\label{fig:EXdyn} 
		Evolution of the kinetic energy as computed from the following dynamics (top to bottom): PAA, GPE, LMA and non-interacting.  The parameters here are $g=10$, $K=12$ and $\kb=2.89$. The agreement between the exact dynamics governed by the GPE and the PAA approximation is very good up to $t=10^3.$ At a longer time, the inset shows significant deviations.}  
	\end{figure}
%-----------------------------------------%

Within the PAA, the GPE evolution of $\hat\psi(p,t)$ between kicks is replaced by
\begin{equation}
i\kb\partial_t\hat\psi =\frac{p^2}{2}\hat\psi +g\dfrac{|F_{\rm PAA}(p,\hat\psi)|}{|\hat\psi|}\hat\psi.
\label{eq:freeGPEPAA}
\end{equation}
The advantage of the PAA is that, as in the LMA, the amplitude of $\hat\psi(p)$ is conserved dynamically between the kicks, implying that $|F_{\rm PAA}(p,\hat\psi)|/|\hat\psi|$ is a constant of motion between kicks, making the integration trivial \footnote{While $|F_{\rm PAA}(p,\hat\psi)|$ is typically broader than $|\hat\psi(p)|$, making $|F_{\rm PAA}(p,\hat\psi)|/|\hat\psi(p)|\gg 1$ for $|\hat\psi(p)|\ll 1$, this is not a problem in the numerics since the phase of $\hat\psi(p)$ does not matter for the dynamics in this regime. },
\begin{equation}
\hat\psi(p,t+1) =\exp\left[-i\left(\frac{p^2}{2} +g\frac{|F_{\rm PAA}(p,\hat\psi)|}{|\hat\psi(p)|}\right)\right]\hat\psi(p,t).
\end{equation}

Fig.~\ref{fig:EXdyn} shows the dynamics obtained from the GPE, LMA, and PAA, as well as the non-interacting case. While the short-time dynamics is interaction-independent (and thus approximation independent), the LMA fails to capture the regime where the interactions start to be relevant, while this is well achieved by the PAA. This improved approximation is a faithful description of the exact GPE dynamics for the typical time range $[0,1000]$ accessible to state-of-the-art experiments (see Sec.~\ref{sec:exp}). In particular, it captures well the timescale at which the interactions start to matter, contrary to the LMA.
As already explained above and illustrated in Fig.~\ref{fig:hist}, at longer, experimentally inaccessible, times, however, both the LMA and the PAA disagree with the exact integration of the GPE, probably due to the nonlinear correlations between the phases which start to build up. Note that we have extracted the subdiffusive exponents of the LMA and PAA, shown in Fig.~\ref{fig:hist}, from the long term dynamics in the typical time-window $t\in[10^4,10^5]$ (data  shown in the inset of Fig.~\ref{fig:EXdyn}). In particular, we recall again that none of these approximations is able to successfully capture the correct subdiffusive exponents for the GPE, as it is clear from Fig.~\ref{fig:hist}.

\section{Can subdiffusion be observed experimentally?}
\label{sec:exp}
Both the interacting kicked rotor and the analogous Anderson model~\cite{Pikovsky:DestructionAndersonLocNonlin:PRL08} are predicted to display a subdiffusive behavior at long times. Subdiffusion has indeed been experimentally 
observed in the closely related Aubry-Andr\'e model with interactions~\cite{Lucioni:SubdiffusionInteract:PRL11}. The main limitations for observing subdiffusion in the nonlinear kicked rotor are threefold: First, quantum dynamics is limited by decoherence effects, like spontaneous emission or stochastic fluctuations of the optical potential; second, free cold atoms fall under gravity and exit the interaction region; third, kicks induce a strong diffusion in real space that very quickly makes the atomic gas so diluted that interactions become negligible.

Decoherence can be reduced by increasing the laser-atom detuning and by carefully crafting the laser beams forming the kicking potential. Recently, an increase from a limit of $\sim$ 200 to 1000 kicks has been obtained, allowing for the observation of the 2D Anderson-like localization~\cite{Manai:Anderson2DKR:PRL15}, and this limit might be increased in the near future. As shown in Fig.~\ref{fig:GPEdyn} subdiffusion can yet be observed at a few hundred kicks for a large enough nonlinearity $g\sim 10$. The use of Feshbach resonances allows one to easily reach even higher values of $g$ [see Eq.~(\ref{eq:g}) and the discussion following it].

It is now possible to create and exploit sophisticated atom traps, in which a Bose-Einstein condensate can be created or inserted. A particularly interesting geometry is a ring trap created by Laguerre-Gauss modes. Such a ring trap can keep atoms for quite long times, allowing for a large number of kicks. Inside the ring, atoms can freely evolve in the azimuthal direction realizing an ``atom rotor'', which corresponds to the original version of the kicked rotor; if the radial confinement energy $\hbar\omega_\perp$ is large compared to the other energy scales, the dynamics is effectively 1D. Moreover, Laguerre-Gauss modes display an azimuthal dependence $\exp(-im\varphi)$ where $\varphi$ is the azimuthal angle. Hence, by combining two modes with azimuthal ``quantum numbers'' $m$ and $-m,$ one creates an azimuthal modulation of the form $\cos(2m\varphi)$ that can be pulsed to transfer angular momentum to the atoms, thus realizing a genuine kicked rotor. Finally, if the ring is initially homogeneously filled with the condensate, no spatial dilution will occur as the kicks are applied. It is worth noting that the ring geometry generates a kicked rotor with true periodic boundary conditions.

We can thus conclude that the observation of subdiffusion is within reach of state-of-the-art experimental setups.

\section{Conclusion}
\label{conclusion}
The study of the exact GPE dynamics has confirmed the breakdown of dynamical localization observed within the LMA but invalidated the universality of the subdiffusion exponent $\alpha$. Moreover, we have introduced a new approximation to the GPE, the PAA, which is better justified than the LMA, yet computationally as advantageous, and allows for a good description of the dynamics up to relatively long times where interactions do play a non-negligible role. However, it does not give the correct range of the subdiffusive exponent. The understanding of the breakdown of the PAA at longer times is left for future work. Explaining the variations of the non-universal exponent $\alpha$ in the exact GPE is also a very interesting line of research.
	
\section*{Acknowledgments}
We thank C. Besse, G. Dujardin and P. Parnaudeau for insightful discussions.
This work was supported by Agence Nationale de la Recherche through Research Grants MANYLOK No. ANR-18-CE30-0017 and QRITiC I-SITE ULNE/ ANR-16-IDEX-0004 ULNE, the Labex CEMPI Grant No.ANR-11-LABX-0007-01, the Programme Investissements d'Avenir ANR-11-IDEX-0002-02, reference ANR-10-LABX-0037-NEXT and the Ministry of Higher Education and Research, Hauts-de-France Council and European Regional
Development Fund (ERDF) through the Contrat de Projets \'Etat-Region (CPER Photonics for Society, P4S). S.L. acknowledges funding by the Labex CEMPI and the co-funded PRESTIGE programme (Campus France, European Union).

\bibliography{artdatabase_v23}

\end{document}